\def\pmb#1{\setbox0=\hbox{$#1$}%
     \kern-.025em\copy0\kern-\wd0
     \kern.05em\copy0\kern-\wd0
     \kern-.025em\raise.0433em\box0 }
\def\spmb#1{\setbox0=\hbox{$\scriptsize #1$}%
     \kern-.020em\copy0\kern-\wd0
     \kern.03em\copy0\kern-\wd0
     \kern-.020em\raise.0433em\box0 }
\def\int{\intop\limits}
\def\bea{\begin{eqnarray}}
\def\eea{\end{eqnarray}}
\def\<{\begin{equation}}
\def\>{\end{equation}}
\def\non{\nonumber}

\def\dst{\displaystyle}

\def\smash#1{\hbox to 0pt{$\dst #1 $\hss}}

\def\nextpage{\bigskip\hrule height2pt
              \ifodd\pagenumber\newpage\phantom{a}\fi
              \cleardoublepage}

\def\log{\mathop{\rm ln}\nolimits}

\def\int{\intop\limits}
\def\bea{\begin{eqnarray}}
\def\eea{\end{eqnarray}}
\def\<{\begin{equation}}
\def\>{\end{equation}}
\def\non{\nonumber}
\def\log{\mathop{\rm ln}\nolimits}

\def\.{\hskip -1pt}

\def\g5{\gamma_5}

\documentstyle[manuscript,aps]{revtex}
\begin{document}
\draft
\title{
Goldstone Theorem in the Gaussian Functional \\
Approximation to the Scalar $\phi^{4}$ Theory}
\author{V.~Dmitra\v sinovi\' c, J.R. Shepard}
\address{ 
   Nuclear Physics Laboratory, Physics Department,\\
   University of Colorado, P.O.Box 0446, 
   Boulder, CO 80309-0446}
\author{J.A. McNeil}
\address{Physics Department,\\
Colorado School of Mines, Golden, CO 80401}
\maketitle
\begin{abstract}
We verify the Goldstone theorem in the Gaussian functional approximation to the $\phi^{4}$ theory with internal O(2) symmetry. We do so by reformulating the Gaussian approximation in terms of Schwinger-Dyson equations from which an explicit demonstration of the Goldstone theorem follows directly. 
\end{abstract}
\widetext
In recent years we have seen a resurgence of interest in a non-perturbative method that is based on the Schr\" odinger picture of quantum field theory (QFT) and goes under the name of Gaussian approximation to the ground state wave functional, due mostly to the work of Barnes and Ghandour \cite{bg80} and J. Kuti (see Sec. V.C. of Ref.\cite{cjt74}). 
Many of these studies have addressed $\phi^{4}$ scalar theories \cite{stev84,cc85}. An explicit relation between the Gaussian functional and the symmetry-conserving Hartree-Fock approximation \cite{fw71} has been established by Cornwall, Jackiw and Tomboulis \cite{cjt74}, and in a somewhat different form by Consoli and Ciancitto \cite{cc85}. However, the Goldstone theorem \cite{gold61}, which is an exact result for $\phi^{4}$ scalar theories with internal symmetry (e.g. O(N) symmetric theories), was apparently shown {\it not} to be satisfied in this approximation \cite{bc85}. We will show in this note that this is not a shortcoming of the Gaussian approximation, but rather a consequence of an incomplete analysis that can readily be remedied. Put another way, the Goldstone theorem is satisfied, but the mechanism is subtle in that it requires an analysis of four-point functions, in close analogy with Nambu's \cite{njl61} proof of Goldstone theorem in self-interacting fermion t!
heories. We emphasize that the pro
of does not depend on the specific values of the bare parameters of the theory, as long as the system is in the spontaneously broken phase. Although the Gaussian functional formalism allows us to map out the parameter space in which the system is in the broken or symmetric phase, we consider it a separate issue that will not be addressed here.

This analysis is most straightforwardly accomplished using Schwinger-Dyson (SD) equations whose form we now discuss. It is known \cite{bg80} that the Gaussian approximation to theories without spontaneous symmetry breaking can be formulated as a specific truncation of the exact Schwinger-Dyson equations \cite{riv87} for the (connected, but one-particle reducible) two- and four-point Green functions. For the purpose at hand we show how to extend this correspondence to the case with spontaneous symmetry breaking \cite{dmit94}. It turns out that a new, one-point Green function SD equation\footnote{This SD equation can be found, although expressed in a slightly different form, in B.W. Lee's work \cite{lee72}.} appears and the two- and four-point Green functions are crucially modified. The validity of the Goldstone theorem is conventionally verified \cite{gold61,lee72} by looking at the two-point Green functions (single-particle propagators) and finding zero-mass poles in them. I!
n the present case, however, we fi
nd that the Goldstone theorem is reflected in the zero-mass poles of the four-point functions (two-body propagators). These we interpret as bound states with all the properties of Goldstone bosons very much like the bound-state Goldstone bosons in the Nambu and Jona-Lasinio (NJL) model \cite{njl61}.

For the sake of simplicity, we follow Brihaye and Consoli \cite{bc85} and confine ourselves to an O(2) symmetric theory. The Lagrangian density is given by
\bea
{\cal L} =  {1 \over 2} \left(\partial_{\mu} {\vec \phi}\right)^{2} - V({\vec \phi}^{2}) ~.
\label{e:lag}\eea
with 
$${\vec \phi} = (\phi_{1},\phi_{2})$$ 
and
$$V({\vec \phi}^{2}) = - {1 \over 2} m_{0}^{2}{\vec \phi}^{2} + {\lambda_{0} \over 4}\left({\vec \phi}^{2}\right)^{2}. $$
We assume here $\lambda_{0}$ and $m_{0}^{2}$ are positive which ensures spontaneous symmetry breaking at the tree level. Unlike Brihaye and Consoli \cite{bc85} who work with ``rotated fields" $(\xi , \eta)$, we will work with ``Cartesian fields" 
\bea
\phi_{1} &=& \langle \phi_{1} \rangle + \xi \cos \theta - \eta \sin \theta \non \\
\phi_{2} &=& \langle \phi_{2} \rangle + \xi \sin \theta + \eta \cos \theta \non \
\eea 
in terms of which the Gaussian ground state functional Ansatz is
\bea
\Psi_{0}[{\vec \phi}] = N \exp \left( - {1 \over 4 \hbar} \int d {\bf x} \int d {\bf y} \left[\phi_{i}({\bf x}) - \langle  \phi_{i}({\bf x}) \rangle \right] 
G_{ij}^{-1}({\bf x},{\bf y})\left[\phi_{j}({\bf y}) - \langle \phi_{j} ({\bf x})) \rangle \right]\right)~,
\label{e:gaus}
\eea
where $N$ is the normalization constant, $\langle  \phi_{i}({\bf x}) \rangle$ is the vacuum expectation value (v.e.v.) of the $i$-th scalar field which henceforth we will assume to be translationally invariant $\langle  \phi_{i}({\bf x}) \rangle = \langle  \phi_{i}(0) \rangle \equiv \langle  \phi_{i} \rangle$ and
$$G_{ij}({\bf x},{\bf y}) = {1 \over 2} \delta_{ij} \int {d {\bf k} \over (2 \pi)^{3}} {1 \over \sqrt{{\bf k}^{2} + m_{i}^{2}}} e^{i {\bf k} \cdot ({\bf x} - {\bf y})} 
.$$ 
Although the present approach lacks {\it manifest} O(2) invariance \footnote{The manifestly O(2) symmetric Gaussian Ansatz $G_{ij}(m_{i}^{2}) = \delta_{ij} G(m^{2})$ is known \cite{cc85} to lead to the symmetric ground state, $m_{i} = m$, and is hence abandoned here.}, its use does not involve any loss of generality as compared with Ref.\cite{bc85}, as can be seen from the resulting equations of motion which are equivalent in the two approaches. 
Furthermore, note that we have explicitly kept $\hbar$ (while setting the velocity of light $c = 1$) to keep track of quantum corrections and count the number of ``loops" in our calculation. Then the vacuum energy density becomes
\bea
{\cal E}(m_{i}, \langle  \phi_{i} \rangle) &=& \hbar I_{1}(m_{1}) + \hbar I_{1}(m_{2}) 
- {\hbar \over 2} \left[m_{1}^{2} I_{0}(m_{1}) + m_{2}^{2} I_{0}(m_{2}) \right] 
\non \\
&-& {1 \over 2} m_{0}^{2} \left[\langle \phi_{1}^{2} \rangle + \langle \phi_{2}^{2} \rangle + \hbar I_{0}(m_{1}) + \hbar I_{0}(m_{2}) \right] + {\lambda_{0} \over 4} \lbrace \left(\langle \phi_{1}^{2} \rangle + \langle \phi_{2}^{2} \rangle \right)^{2}  \non \\
&+&\langle \phi_{1}^{2} \rangle  \hbar\left[6 I_{0}(m_{1}) + 2 I_{0}(m_{2}) \right] + \langle \phi_{2}^{2} \rangle \hbar \left[6 I_{0}(m_{2}) + 2 I_{0}(m_{1}) \right] \non \\
&+&  3 \left(\hbar I_{0}(m_{1})\right)^{2} + 3 \left(\hbar I_{0}(m_{2})\right)^{2} + 2 \hbar^{2} I_{0}(m_{1}) I_{0}(m_{2})  \rbrace \ ~, 
\label{e:ener}
\eea
where 
\bea
 I_{0}(m_{i}) &=& {1 \over 2} \int {d {\bf k} \over (2 \pi)^{3}} {1 \over \sqrt{
{\bf k}^{2} + m_{i}^{2}}} = i \int {d^{4} k \over (2 \pi)^{4}} {1 \over {k^{2} - m_{i}^{2} + i \varepsilon}} = G_{ii}({\bf x},{\bf x}) \non \\
 I_{1}(m_{i}) &=& {1 \over 2} \int {d {\bf k} \over (2 \pi)^{3}} \sqrt{{\bf k}^{2} + m_{i}^{2}} = - {i \over 2} \int {d^{4} k \over (2 \pi)^{4}} \log \left(k^{2} - m_{i}^{2} + i \varepsilon \right) ~.\non \
\eea
The divergent integrals $I_{0,1}(m_{i})$ are understood to be regularized via an UV momentum cut-off. We may identify $\hbar I_{1}(m_{i})$ as the familiar ``zero-point" energy density of a free scalar field of mass $m_{i}$. 

We vary the energy density with respect to the field vacuum expectation values $\langle \phi_{i} \rangle$ and the ``dressed" masses $m_{i}$. The results allow a straightforward interpretation in terms of Feynman diagrams, i.e. truncated  SD equations. 
The extremization condition with respect to the field vacuum expectation values reads:
$$\left({\partial {\cal E}(m_{i}, \langle  \phi_{i} \rangle) \over \partial \langle \phi_{i} \rangle}\right)_{min} = 0, ~~~i = 1,2~, $$
or
\bea
\left({\partial {\cal E}(m_{i}, \langle  \phi_{i} \rangle ) \over{\partial\langle  \phi_{1} \rangle}}\right)_{min} &=& \langle  \phi_{1} \rangle \left[- m_{0}^{2} +  {\lambda_{0}} \left(\langle \phi_{1}^{2} \rangle + \langle \phi_{2}^{2} \rangle + 3 \hbar I_{0}(m_{1}) + \hbar I_{0}(m_{2}) \right) \right]_{min} = 0 \non \\
\left({\partial {\cal E}(m_{i}, \langle  \phi_{i} \rangle ) \over{\partial\langle \phi_{2} \rangle}}\right)_{min} &=& \langle  \phi_{2} \rangle \left[- m_{0}^{2} +  {\lambda_{0}} \left(\langle \phi_{1}^{2} \rangle + \langle \phi_{2}^{2} \rangle + \hbar I_{0}(m_{1}) + 3 \hbar I_{0}(m_{2}) \right) \right]_{min} = 0 ~.
\label{e:vac} \eea
Note that if we assume that both $\langle \phi_{1} \rangle$ and $\langle \phi_{2} \rangle$ are simultaneously nonzero, then after subtracting one of the Eqs. (\ref{e:vac}) from the other, we are forced to conclude that $I_{0}(m_{1}) = I_{0}(m_{2}) $. This is equivalent to the statement that the two masses squared are identical, unless the inverse of $I_{0}(m_{i})$ is a multi-valued function. That, of course, leads us straight to the symmetric phase of the theory, so we discard this possibility. Instead we assume only one $\langle  \phi_{i} \rangle$ to be nonzero, say $\langle \phi_{1} \rangle \neq 0$, while $\langle  \phi_{2} \rangle = 0$. The resulting equations 
\bea
m_{0}^{2} &=& {\lambda_{0}} \left(\langle \phi_{1}^{2} \rangle + 3 \hbar I_{0}(m_{1}) + \hbar I_{0}(m_{2}) \right) \non \\
\langle  \phi_{2} \rangle  &=& 0 \ ~.
\label{e:vev} \eea
can be identified as the truncated SD equation \cite{dmit94} for the one-point Green function (the scalar field v.e.v.) depicted in Fig. 1. with correct symmetry numbers of contributing Feynman diagrams automatically included. As in the ordinary perturbative treatment of the linear sigma model, we associate the nonvanishing v.e.v. with the ``sigma meson" field, and the second field will be called the ``pion". Note that to ${\cal O}(\hbar^{0})$, Eq. (\ref{e:vev}) coincides with the classical result, as it should. Next extremize with respect to the ``dressed" masses $m_{i}$:
$${\partial {\cal E}(m_{i}, \langle  \phi_{i} \rangle) \over \partial m_{i}} = 0, ~~~i = 1,2~, $$
or
\bea
\left({\partial {\cal E}(m_{i}, \langle  \phi_{i} \rangle ) \over{\partial m_{1}}}\right)_{min} &=& \hbar{\partial I_{0} (m_{1}) \over{\partial m_{1}}} \left[ - m_{1}^{2} - m_{0}^{2} +   {\lambda_{0}} \left(3 \langle \phi_{1}^{2} \rangle + \langle \phi_{2}^{2} \rangle  + 3 \hbar I_{0}(m_{1}) + \hbar I_{0}(m_{2}) \right) \right]_{min} = 0 \non \\
\left({\partial {\cal E}(m_{i}, \langle  \phi_{i} \rangle ) \over{\partial m_{2}}}\right)_{min} &=& \hbar{\partial I_{0} (m_{2}) \over{\partial m_{2}}} \left[ - m_{2}^{2} - m_{0}^{2} +  {\lambda_{0}} \left(\langle \phi_{1}^{2} \rangle + 3 \langle \phi_{2}^{2} \rangle  + \hbar I_{0}(m_{1}) + 3 \hbar I_{0}(m_{2}) \right) \right]_{min}= 0 \ ~.
\label{e:gap}
\eea
Eqs.(\ref{e:gap}) also have a Feynman-diagrammatic interpretation shown in Fig. 2. that is recognized as a truncated SD equation\cite{dmit94}. Insert Eq. (\ref{e:vev}) into Eq. (\ref{e:gap}) above and find the following two ``gap" equations
\begin{mathletters}
\bea
 m_{1}^{2} &=& 2 {\lambda_{0}} \langle \phi_{1}^{2} \rangle = 2 \left[ m_{0}^{2} - {\lambda_{0}} \hbar \left(3 I_{0}(m_{1}) + I_{0}(m_{2})\right)\right] 
\label{e:gap1a}\\
 m_{2}^{2} &=& 2 {\lambda_{0}} \hbar \left(I_{0}(m_{2}) - I_{0}(m_{1})\right) \ ~.
\label{e:gap1b}
\eea
\end{mathletters}
in agreement with Brihaye and Consoli \cite{bc85}. The system of Eqs. (\ref{e:gap1a},b) admits only massive solutions $m_{1} > m_{2} > 0$ for real, positive values of $m_{0}^{2}, \lambda_{0}$ and a finite ultraviolet cut-off of the momentum integrals $I_{0,1}m_{i}$. We see that at the tree level, ${\cal O}(\hbar^{0})$, the ``pion" is massless, but the non-perturbative one-loop corrections, ${\cal O}(\hbar)$, give it a finite mass. On the basis of Eq. (\ref{e:gap1b}), it was concluded \cite{bc85} that the Gaussian approximation does not satisfy the Goldstone theorem\footnote{There is another claim \cite{rk87} of failure of ``Goldstone phenomena" in the literature; it, however, does not concern the existence of the Goldstone boson in the broken phase, but rather the criteria for the presence of the broken phase of the theory.}. It is our contention that this does not prove the breakdown of the approximation. Indeed, we show that the Goldstone particle appears as a pole in the !
{\it two-particle propagator} i.e.
 it is a bound state of the two distinct massive elementary excitations in the theory. In order to prove this statement, we must specify the two-body dynamics in the theory. That will be done in terms of the four-point SD or, equivalently, the Bethe-Salpeter, equation.

As mentioned above and noted by Barnes and Ghandour \cite{bg80}, Eqs. (\ref{e:gap1a},b) have an interpretation in terms of a truncated two-point SD equation, as shown diagrammatically in Fig. 2. 
The difference between our Fig. 2. and Fig. 1. of Ref. \cite{bg80} is the presence of the ``tadpole" diagram in the former. That is a consequence of the spontaneous symmetry breaking in the present case. This shows that the Nambu-Goldstone mode of our theory has substantially modified the two-point SD equation from its Wigner-Weyl mode counterpart. 
It will, then, come as no surprise that other, higher-order, SD equations are also crucially modified, the four-point Green function being just one of them.
We write down the correct form of the four-point SD equation and leave the derivation for a longer paper \cite{dmit94}. Specifically, the broken-symmetry four-point Green function SD equation for the scattering of two non-identical (``pion-sigma" scattering) scalar particles has the same generic form of a geometric series as in the symmetric phase, shown in Fig. 2. of Ref. \cite{bg80}, but with a different iterated ``kernel" $\Pi $. The kernel $\Pi $ in the broken phase contains a ``tadpole" diagram (Fig. 3) not present in the symmetric phase. Note that only the $s-$ and $u-$channels carry the ``isospin" quantum numbers of the ``pion", and this is where we look for Goldstone particles. The $t-$channel exchanged ``resonance" carries quantum numbers of the ``sigma meson", and is hence omitted in what follows. For definiteness, focus on the $s$-channel part $T(s)$ of the total four-point scattering amplitude $T(s,t,u)$; its four-point SD equation reads 
\bea
i T(s) = - 2 i M(s) =  - 2 i \lambda_{0} + 2 i M(s) \Pi(s)
\label{e:bse}
\eea
with the simple solution
$$M(s) =  {\lambda_{0} \over{1 - \Pi(s)}}$$
where $s = (p_{1} + p_{2})^{2} \equiv P^{2}$ is the total center-of-mass (CM) energy. A massless bound state in the $s-$channel is manifested in the vanishing of the denominator at $s = 0$. The kernel (``polarization function") $\Pi(s)$ can be read off from the diagrams in Fig. 3. using the Feynman rules for the Lagrangian Eq. (\ref{e:lag}) given in Ref.\cite{lee72}:
\bea
\Pi(s) &=& \Pi_{1}(s) + \Pi_{2}(s) \non \\
\Pi_{1}(s) &=& - 2 i {\lambda_{0}} \hbar \int {d^{4} k \over (2 \pi)^{4}} 
{i^{2} \over {\left[k^{2} - m_{1}^{2} + i \varepsilon \right]\left[(k - P)^{2} - m_{2}^{2} + i \varepsilon \right]}}  \non \\
\Pi_{2}(s) &=& \left(- 2 i {\lambda_{0}} \langle \phi_{1} \rangle \right)^{2}  
\left({i \over s - m_{2}^{2}}\right) \hbar 
\int {d^{4} k \over (2 \pi)^{4}} {i^{2} \over {\left[k^{2} - m_{1}^{2} + 
i \varepsilon \right]\left[(k - P)^{2} - m_{2}^{2} + i \varepsilon \right]}}  \ ~~.
\label{e:pol}
\eea
Next we look at the zero CM energy $P = 0$ polarization function:
\bea
\Pi_{1}(0) &=& {2 {\lambda_{0}} \hbar \over (m_{1}^{2} - m_{2}^{2})} \left[I_{0}(m_{1}) - I_{0}(m_{2})\right]  \non \\
\Pi_{2}(0) &=& {\left(2 {\lambda_{0}} \langle \phi_{1} \rangle \right)^{2} \over - m_{2}^{2}}\left({\hbar \over m_{1}^{2} - m_{2}^{2}}\right) \left[I_{0}(m_{1}) - I_{0}(m_{2})\right]  \
~.
\label{e:polz}
\eea
Now use Eq. (\ref{e:gap}b) to write
\bea
\Pi(0)
 &=& {2 {\lambda_{0}} \hbar \over (m_{1}^{2} - m_{2}^{2})} \left[I_{0}(m_{1}) - I_{0}(m_{2})\right] \left[1 - {2 {\lambda_{0}} \langle \phi_{1}^{2} \rangle \over m_{2}^{2}}  \right] \non \\  
&=& {- m_{2}^{2} \over (m_{1}^{2} - m_{2}^{2})} \left[1 - {2 {\lambda_{0}} \langle \phi_{1}^{2} \rangle \over m_{2}^{2}}  \right] \
~
\label{e:polz2}
\eea
and then use Eq. (\ref{e:gap}a) to obtain the final result
\bea
\Pi(0)
 &=& {- m_{2}^{2} \over (m_{1}^{2} - m_{2}^{2})} \left[1 - {m_{1}^{2} \over m_{2}^{2}}  \right] = 1 \
~.
\label{e:polz3}
\eea
Eq. (\ref{e:polz3}) demonstrates the existence of a Goldstone pole in the two-particle scattering amplitude. This bound state is massless despite the fact that the constituents are massive. This is entirely analogous to the results found in the NJL model \cite{njl61}. The main difference between the present model and the NJL one is that the elementary excitations are bosons instead of fermions in the latter. Note that Eq. (\ref{e:polz3}) is a consequence of all three SD Eqs. (\ref{e:vev},\ref{e:gap1a},b,\ref{e:bse}) discussed here, which are also coupled among themselves in the sense that the solution to one enters the definition of the other. This means that this is a closed, albeit truncated, self-consistent nonperturbative approximation of the kind that is well known under the name of Hartree-Fock plus Random-Phase Approximation in other areas of quantum physics\cite{fw71}. Our main contribution is the symmetry-preserving extension of this scheme to the case of boson fiel!
d theory with spontaneously broken
 symmetry.

The mode of spontaneous symmetry breaking discussed above is something of a surprise: normally, e.g. to any finite order in perturbation theory, we would have expected the v.e.v.s of the two scalar fields to lie on the ``chiral circle", the field with vanishing v.e.v. being massless. The breaking of this ``traditional paradigm" can be traced back to the presence of ${\cal O}(\hbar^{2})$ terms in Eq. \ref{e:ener}. We cannot, however, discard those terms with impunity because in doing so we would upset the delicate self-consistency of this closed set of SD equations. We have no explanation for this unexpected phenomenon, except the suggestion that perhaps it is the bound states that ought to lie on the chiral circle, and not the elementary excitations, again in analogy with the NJL model. 
 In conclusion, we have shown that the Gaussian functional approximation is a self-consistent non-perturbative approximation to the O(2) symmetric linear sigma model that preserves Goldstone's theorem albeit in a subtle way.

This work was supported by the US DOE.

\widetext
\begin{figure}

\caption{Feynman diagrams describing the one-point Green function Schwinger-Dyson equation: the one-loop graph (a), and the tree tadpole diagram (b). The solid line denotes the bare meson doublet, double solid line is the dressed meson doublet. The shaded blob together with the double line leading to it (the ``tadpole") denotes the vacuum expectation value of the field (i.e. the one-point Green function) and the solid dot in the intersection of four lines denotes the bare four-point coupling. Diagrams are explicitly multiplied by their symmetry numbers.}
\label{1p}

\caption{Feynman diagrams describing the two-point Green function Schwinger-Dyson equation: the one-loop graph (a), the tree tadpole diagram (b). The symbols have the same meaning as in Fig. 1.}
\label{2p}

\caption{Feynman diagrams describing the kernel $\Pi (s)$ entering the four-point Green function Schwinger-Dyson equation: the old bubble graph (a), and the new tadpole diagram (b). The symbols have the same meaning as in Fig. 1., but the dot on the right-hand side of the diagrams does not have a $\lambda_{0}$ associated with it, rather it only denotes the separation point of two propagators.}
\label{3p}

\label{disp}

\end{figure}
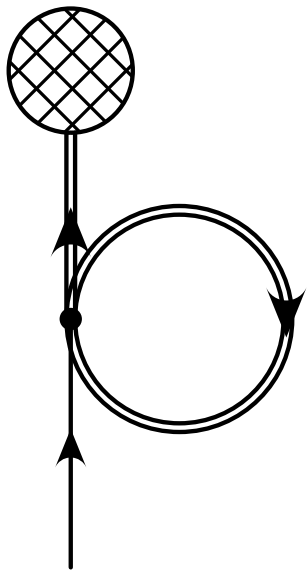
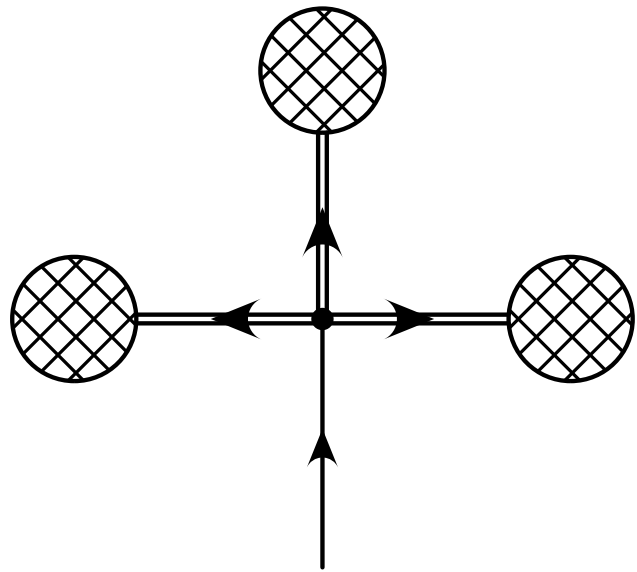
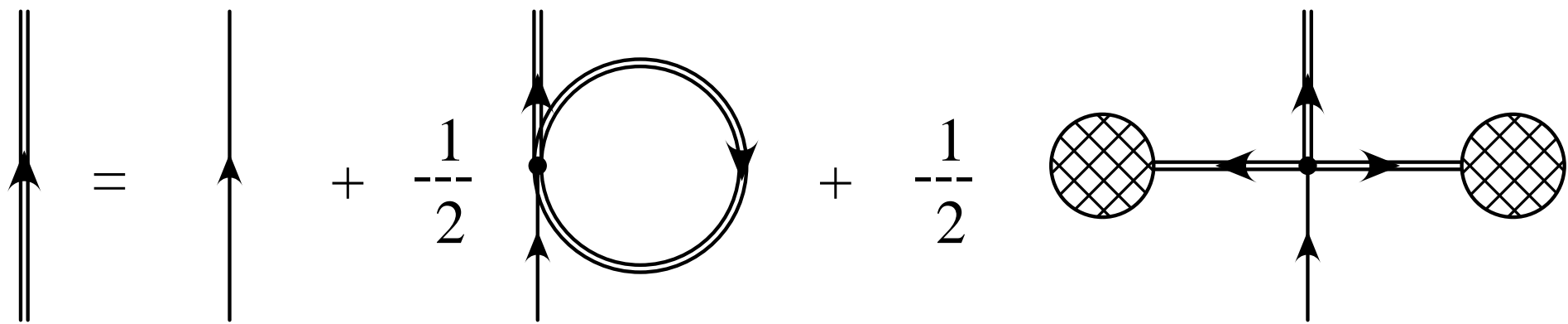
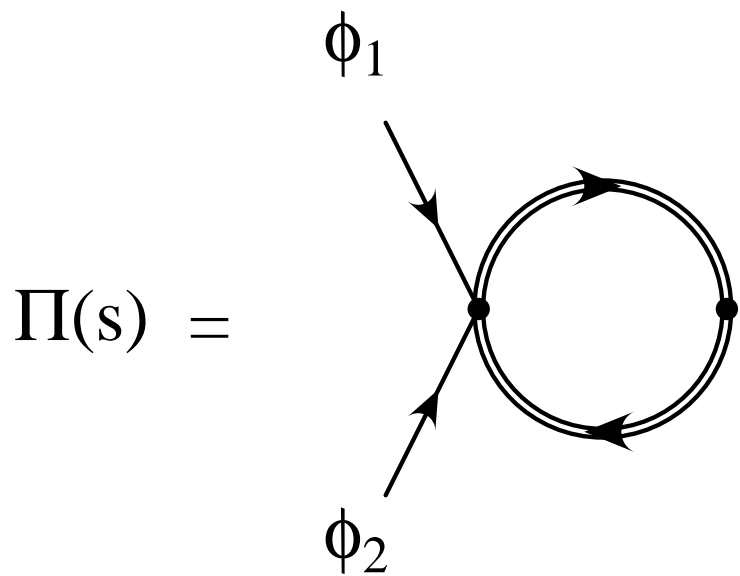
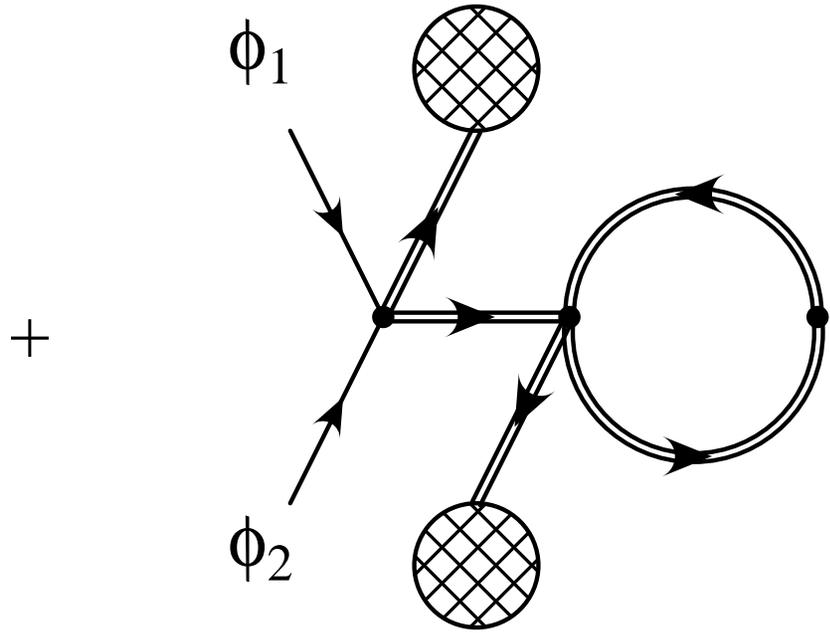
\end{document}